\documentclass[11pt,preprintnumbers,titlepage]{revtex4-2}
\usepackage[latin9]{inputenc}
\usepackage{amsmath}
\usepackage{amssymb}
\usepackage{graphicx}
\usepackage[unicode=true,
 bookmarks=false,
 breaklinks=false,pdfborder={0 0 1},backref=section,colorlinks=false]
 {hyperref}
\hypersetup{
 colorlinks,linkcolor=blue,citecolor=blue,urlcolor=blue}

\makeatletter

\usepackage{wasysym}
\usepackage{amsfonts}
\usepackage{subfigure}
\usepackage{cleveref}
\usepackage{listings}
\usepackage{xcolor}
\usepackage{array}
\usepackage{url}
\usepackage{color}
\usepackage{pifont}

\@ifundefined{textcolor}{}{
\definecolor{BLACK}{gray}{0}
\definecolor{WHITE}{gray}{1}
\definecolor{RED}{rgb}{1,0,0}
\definecolor{GREEN}{rgb}{0,1,0}
\definecolor{BLUE}{rgb}{0,0,1}
\definecolor{CYAN}{cmyk}{1,0,0,0}
\definecolor{MAGENTA}{cmyk}{0,1,0,0}
\definecolor{YELLOW}{cmyk}{0,0,1,0}
}

\makeatother

\begin{document}
\preprint{CTP-SCU/2022015}
\title{Black Holes with Multiple Photon Spheres}
\author{Guangzhou Guo$^{a}$}
\email{gzguo@stu.scu.edu.cn}

\author{Yuhang Lu$^{a}$}
\email{luyuhang668@stu.scu.edu.cn}

\author{Peng Wang$^{a}$}
\email{pengw@scu.edu.cn}

\author{Houwen Wu$^{a,b}$}
\email{hw598@damtp.cam.ac.uk}

\author{Haitang Yang$^{a}$}
\email{hyanga@scu.edu.cn}

\affiliation{$^{a}$Center for Theoretical Physics, College of Physics, Sichuan
University, Chengdu, 610064, China}
\affiliation{$^{b}$Department of Applied Mathematics and Theoretical Physics,
University of Cambridge, Wilberforce Road, Cambridge, CB3 0WA, UK}
\begin{abstract}
Recently, asymptotically-flat black holes with multiple photon spheres
have been discovered and found to produce distinctive observational
signatures. In this paper, we focus on whether these black hole solutions
are physically viable, e.g., satisfying energy conditions of interest.
Intriguingly, black hole and naked singularity solutions with two
photon spheres and one anti-photon sphere\ are shown to exist in
physically reasonable models, which satisfy the null, weak, dominant
and strong energy conditions. Our findings reveal that black holes
with multiple photon spheres may not be frequent, but they are not
exotic.
\end{abstract}
\maketitle
\tableofcontents{}

\section{Introduction}

The images of the supermassive black holes M87{*} \cite{Akiyama:2019cqa,Akiyama:2019brx,Akiyama:2019sww,Akiyama:2019bqs,Akiyama:2019fyp,Akiyama:2019eap,Akiyama:2021qum,Akiyama:2021tfw}
and Sgr A{*} \cite{EventHorizonTelescope:2022xnr,EventHorizonTelescope:2022vjs,EventHorizonTelescope:2022wok,EventHorizonTelescope:2022exc,EventHorizonTelescope:2022urf,EventHorizonTelescope:2022xqj}
photographed by the Event Horizon Telescope collaboration and the
gravitational waves from a binary black hole merger detected by LIGO
and Virgo \cite{Abbott:2016blz} will usher in a new era of black
hole observations. For spherically symmetric black holes, photon spheres
play a crucial role in imaging black holes, and generate the edge
of black hole shadows. Moreover, photon spheres are closely related
to quasinormal modes \cite{Cardoso:2008bp}, whose superposition describes
the ringdown stage of a binary black hole merger. Therefore, it is
of great importance to investigate the role played by photon spheres
in black hole observations.

Until recently, asymptotically-flat black holes were supposed to possess
a single photon sphere outside the event horizon, particularly in
a physically reasonable model. The single photon sphere would lead
to a bright ring in black hole images and the absence of echoes in
gravitational waveforms of ringdown signals. Intriguingly, the existence
of two photon spheres outside the event horizon has been lately reported
for dyonic black holes in Einstein's gravity where the quasi-topological
electromagnetism is minimally coupled \cite{Liu:2019rib}. Later,
it was found that the double-peak effective potential of scalar perturbations
in the dyonic black holes with double photon spheres can give rise
to echo signals \cite{Huang:2021qwe}.

Meanwhile, scalarized Reissner-Nordström (RN) black holes were constructed
to understand the formation of hairy black holes \cite{Herdeiro:2018wub},
and we found that scalarized RN black holes can possess two photon
spheres outside the event horizon in certain parameter regions \cite{Gan:2021pwu}.
Subsequently, optical appearances of accretion disks and luminous
celestial spheres surrounding scalarized RN black holes were considered,
which showed that an extra photon sphere can produce more bright rings
in black hole images, noticeably increase the image flux and triple
higher-order relativistic images \cite{Gan:2021xdl,Guo:2022muy}.
In addition, observational appearances of a star freely falling onto
scalarized RN black holes have been numerically simulated, and it
revealed that an extra photon sphere leads to a sharp peak of total
luminosity and one more cascade of flashes seen by a specific observer
\cite{Chen:2022qrw}.

On the other hand, the authors of \cite{Cvetic:2016bxi} considered
asymptotically-flat and spherically symmetric black hole solutions
in various theories and found that there is only one photon sphere
outside the event horizon if the Dominant Energy Condition (DEC) and
Strong Energy Condition (SEC) are satisfied. This observation then
led to the conjecture, which states that a violation of either SEC
or DEC is a necessary condition for the existence of double photon
spheres outside the event horizon. Additionally, it was proved in
\cite{Cvetic:2016bxi} that SEC requires a monotonically increasing
$\left\vert g_{tt}\right\vert $ in Einstein's gravity. For dyonic
black holes with double photon spheres, $\left\vert g_{tt}\right\vert $
was shown to have a wiggle outside the event horizon, and hence SEC
is violated \cite{Liu:2019rib}. This discovery is consistent with
the conjecture proposed in \cite{Cvetic:2016bxi}.\ Although the
existence of double photon sphere in black hole spacetime is new,
it is well known that wormholes can have two photon spheres, one on
each side of the wormholes. In Einstein's gravity, exotic matter is
usually required to keep the mouths of traversable wormholes open,
which may reinforce the conjecture.

Nevertheless, no fundamental principles prevent us from searching
for black hole solutions with multiple photon spheres in a physically
reasonable model, e.g., satisfying SEC and DEC. To this end, we investigate
various parametrized black holes and specific models to discover solutions
with double photon spheres and check energy conditions for them in
this paper. The rest of the paper is organized as follows. In Section
\ref{sec:EC}, we briefly review four energy conditions and a necessary
condition for SEC. Black hole solutions with double photon spheres
are considered in the context of parametrized black holes and specific
black hole models in Section \ref{sec:DPSBH}. In Section \ref{sec:NLEDBH},
the existence of double photon spheres is discussed in the spacetime
sourced by non-linear electromagnetic fields, where photons move along
null geodesics of some effective metric. Finally, we conclude with
a brief discussion in Section \ref{Sec:Con}. We set $16\pi G=1$
throughout this paper.

\section{Energy Conditions}

\label{sec:EC}

In this section, we briefly review energy conditions in general relativity
and how the strong energy condition constrains black hole metric.
The energy conditions are expected to impose restrictions on the stress-energy-momentum
tensor $T_{\mu\nu}$ of matter fields in a physically reasonable model
\cite{Maeda:2018hqu}. In particular, standard energy conditions for
$T_{\mu\nu}$ in a $D$-dimensional spacetime are stated as follows:

\begin{itemize}
\item Null Energy Condition (NEC): $T_{\mu\nu}l^{\mu}l^{\nu}\geq0$ for
any null vector $l^{\mu}$. It guarantees that the mass-energy density
measured by an observer traversing any null orbit is non-negative.
\item Weak Energy Condition (WEC): $T_{\mu\nu}v^{\mu}v^{\nu}\geq0$ for
any timelike vector $v^{\mu}$. It guarantees that the mass-energy
density measured by an observer traversing any timelike orbit is non-negative.
\item DEC: $T_{\mu\nu}v^{\mu}v^{\nu}\geq0$ and $J^{\mu}J_{\mu}\leq0$ for
any timelike vector $v^{\mu}$, where $J^{\mu}\equiv T_{\nu}^{\mu}v^{\nu}$
is the energy flux measured by an observer with four-velocity $v^{\mu}$.
It guarantees that, for every timelike observer, the measured mass-energy
is non-negative, and the total flux of energy momentum does not propagate
faster than the speed of light.
\item SEC: $\left(T_{\mu\nu}-\frac{Tg_{\mu\nu}}{D-2}\right)v^{\mu}v^{\nu}\geq0$
for any timelike vector $v^{\mu}$. It guarantees that timelike geodesic
congruence locally converges, which implies that ``gravity acts attractively''\ on
matter moving along timelike geodesics.
\end{itemize}

In some orthonormal basis $\left\{ e_{a}^{\mu}\right\} $ satisfying
$e_{a}^{\mu}e_{b\mu}=\eta_{ab}$ and $g^{\mu\nu}=\eta^{ab}e_{a}^{\mu}e_{b}^{\nu}$,
the components of the energy-momentum tensor in the orthonormal frame
$T_{ab}=T_{\mu\nu}e_{a}^{\mu}e_{b}^{\mu}$ can be written as in a
canonical form \cite{Hawking:1973uf}. If the canonical form of $T_{ab}$
is type I (i.e., $T_{ab}=$diag$\left(\rho,p_{1},p_{2},\cdots,p_{D-1}\right)$),
the aforementioned energy conditions can be expressed as follows:

\begin{itemize}
\item NEC: $\rho+p_{i}\geq0$ for $i=1,2,\cdots,D-1.$
\item WEC: $\rho\geq0$ and $\rho+p_{i}\geq0$.
\item DEC: $\rho\geq\left\vert p_{i}\right\vert $ for $i=1,2,\cdots,D-1$
and $\rho\geq0$.
\item SEC: $\left(D-3\right)\rho+\sum\limits _{i=1}^{D-1}p_{i}\geq0$ and
$\rho+p_{i}\geq0$ for $i=1,2,\cdots,D-1$.
\end{itemize}

For a spherically symmetric and static black hole solution in general
relativity
\begin{equation}
ds^{2}=-f\left(r\right)dt^{2}+\frac{dr^{2}}{h\left(r\right)}+R\left(r\right)\left(d\theta^{2}+\sin^{2}\theta d\varphi^{2}\right),\label{eq:SMetric}
\end{equation}
the Einstein field equations yield \cite{Cvetic:2016bxi}
\begin{equation}
f^{\prime}\left(r\right)>\frac{1}{R\left(r\right)}\sqrt{\frac{f\left(r\right)}{h\left(r\right)}}\int_{r_{h}}^{r}R\left(r^{\prime}\right)\sqrt{\frac{h\left(r^{\prime}\right)}{f\left(r^{\prime}\right)}}\left(\rho+p_{1}+p_{2}+p_{3}\right)dr^{\prime}.\label{eq:f(r)SEC}
\end{equation}
Here, $r_{h}$ is the event horizon radius, and the orthonormal frame
is chosen as the static frame,
\begin{equation}
e_{0}^{\mu}dx_{\mu}=\frac{dt}{\sqrt{f\left(r\right)}},e_{1}^{\mu}dx_{\mu}=\sqrt{h\left(r\right)}dr,e_{2}^{\mu}dx_{\mu}=\frac{1}{\sqrt{R\left(r\right)}},e_{3}^{\mu}dx_{\mu}=\frac{1}{\sqrt{R\left(r\right)}\sin\theta}.
\end{equation}
Interestingly, eqn. $\left(\ref{eq:f(r)SEC}\right)$ shows that $f^{\prime}\left(r\right)>0$
is a necessary condition for SEC. In other words, if SEC is respected,
$f\left(r\right)$ must be monotonically increasing. From this property,
it was conjectured in \cite{Cvetic:2016bxi} that, for an asymptotically-flat
black hole, ``a violation of either the dominant or the strong energy
condition is a necessary condition for the existence of an anti-photon
sphere outside a regular black hole horizon.''

\section{Double-Photon-Spheres Black Holes}

\label{sec:DPSBH}

For the black hole solution $\left(\ref{eq:SMetric}\right)$, the
Lagrangian governing null geodesics is
\begin{equation}
\mathcal{L}=\frac{1}{2}\left[-f\left(r\right)\dot{t}^{2}+\frac{\dot{r}^{2}}{h(r)}+R\left(r\right)\left(\dot{\theta}^{2}+\sin^{2}\theta\dot{\varphi}^{2}\right)\right],\label{eq:Geo Lagrangian}
\end{equation}
where dots stand for derivative with respect to the affine parameter
$\lambda$. Due to spherical symmetry, we can confine ourselves to
null geodesics on the equatorial plane. Since $t$ and $\varphi$
do not explicitly appear in the Lagrangian, null geodesics can be
characterized by the energy $E$ and the angular momentum $L$,
\begin{equation}
E=-p_{t}=f\left(r\right)\dot{t}\text{, }L=p_{\varphi}=R\left(r\right)\dot{\varphi}.\label{eq:E-L}
\end{equation}
Moreover, the condition $\mathcal{L}=0$ gives the radial component
of the null geodesic equations,
\begin{equation}
\frac{f\left(r\right)}{h(r)}\frac{\dot{r}^{2}}{L^{2}}+V_{\text{eff}}(r)=\frac{1}{b^{2}},\label{eq:Pr}
\end{equation}
where $b\equiv L/E$ is the impact parameter. Here, we define the
effective potential for null geodesics,
\begin{equation}
V_{\text{eff}}\left(r\right)=\frac{f\left(r\right)}{R\left(r\right)},
\end{equation}
which does not depend on the rr-component of the metric.

A circular null geodesic occurs at an extremum of the effective potential
$V_{\text{eff}}(r)$, and the radius $r_{c}$ of the geodesic is determined
by
\begin{equation}
V_{\text{eff}}\left(r_{c}\right)=b_{c}^{-2}\text{, }V_{\text{eff}}^{\prime}\left(r_{c}\right)=0,
\end{equation}
where $b_{c}$ is the impact parameter of the geodesic. In addition,
local maxima and minima of the effective potential correspond to unstable
and stable circular null geodesics, respectively. The unstable and
stable circular null geodesics would constitute a photon sphere and
an anti-photon sphere, respectively. Moreover, an asymptotically-flat
black hole has $V_{\text{eff}}\left(\infty\right)=0=V_{\text{eff}}\left(r_{h}\right)$
and $V_{\text{eff}}\left(r\right)>0$ for $r>r_{h}$, which indicates
that there must exist at least one photon sphere outside the event
horizon, and the number of photon spheres is always one more than
the number of anti-photon spheres. Note that the same result has been
reported in \cite{Cunha:2020azh,Wei:2020rbh,Ghosh:2021txu,Qiao:2022jlu}.
Particularly, the existence of an anti-photon sphere means two photon
spheres outside the event horizon. If a null stable circular geodesic
is at $r=r_{c}$, the conditions $V_{\text{eff}}^{\prime}\left(r_{c}\right)=0$
and $V_{\text{eff}}^{\prime\prime}\left(r_{c}\right)>0$ give $f^{\prime\prime}\left(r_{c}\right)>0$.
So a black hole with double photon spheres should have $f^{\prime\prime}\left(r\right)>0$
at some point. It is worth noting that NEC has been shown to be satisfied
at anti-photon spheres \cite{Cunha:2017qtt}.

\subsection{Parametrized Black Holes}

In this subsection, we explore parameter regions of the existence
of double photon spheres in some parametrized black hole metrics,
which contain a set of parameters measuring deviations from the Schwarzschild\ metric.
As previously stated, the effective potential $V_{\text{eff}}(r)$
and the necessary condition for SEC (i.e., $f^{\prime}\left(r\right)>0$)
are independent of $h(r)$ while the validity of the energy conditions
generally depends on $h(r)$. Therefore, we also investigate two kinds
of parameter regions where $f^{\prime}\left(r\right)>0$ regardless
of $h(r)$, and the energy conditions are satisfied assuming $f(r)=h(r)$,
respectively.

\subsubsection{Perturbed Schwarzschild Black Holes}

\begin{figure}[t]
\centering %
\begin{minipage}[c]{0.45\linewidth}%
 \subfigure{\includegraphics[scale=0.63]{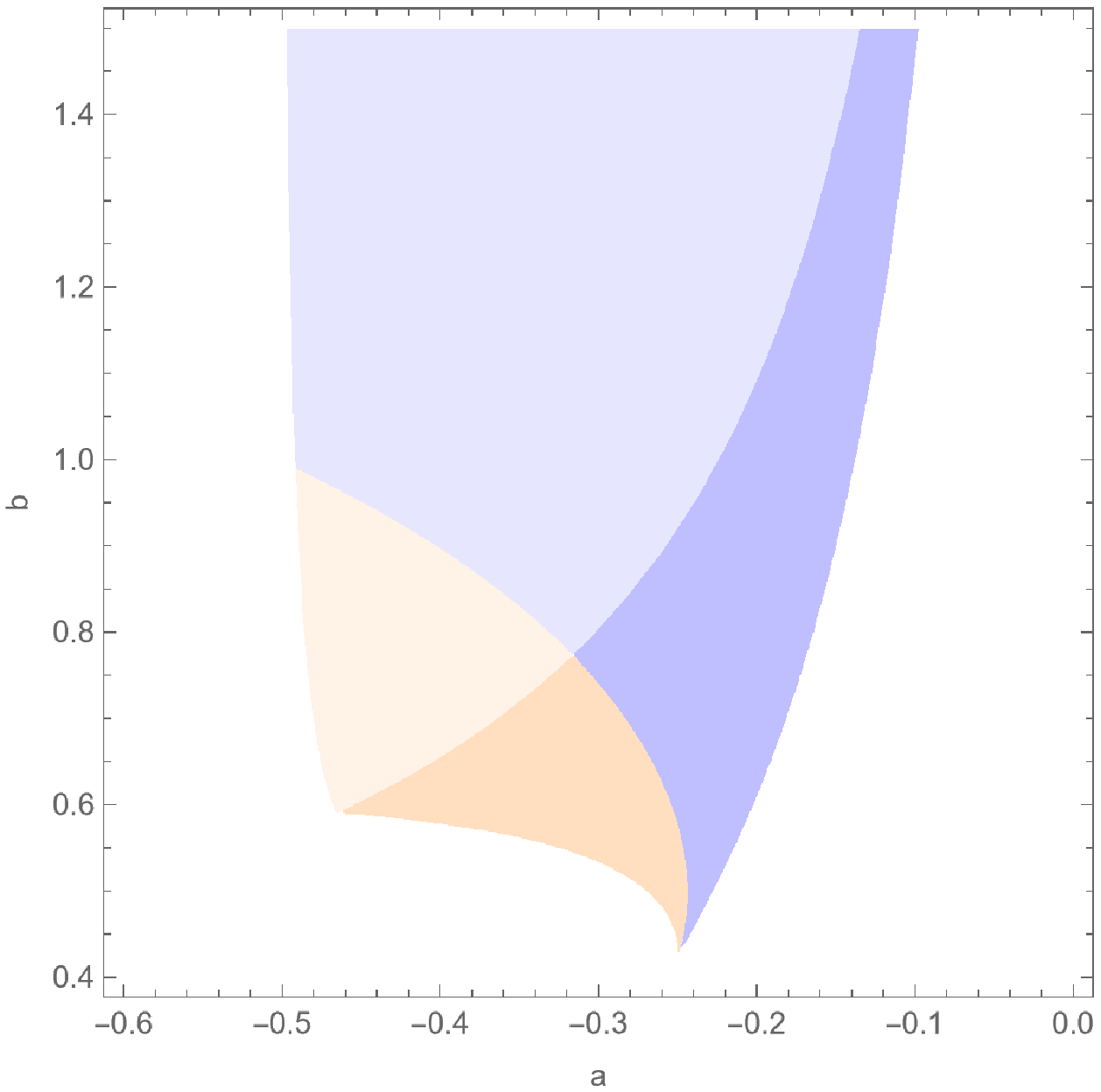}} %
\end{minipage}\hspace{2pt} %
\begin{minipage}[c]{0.45\linewidth}%
 \subfigure{\includegraphics[scale=0.48]{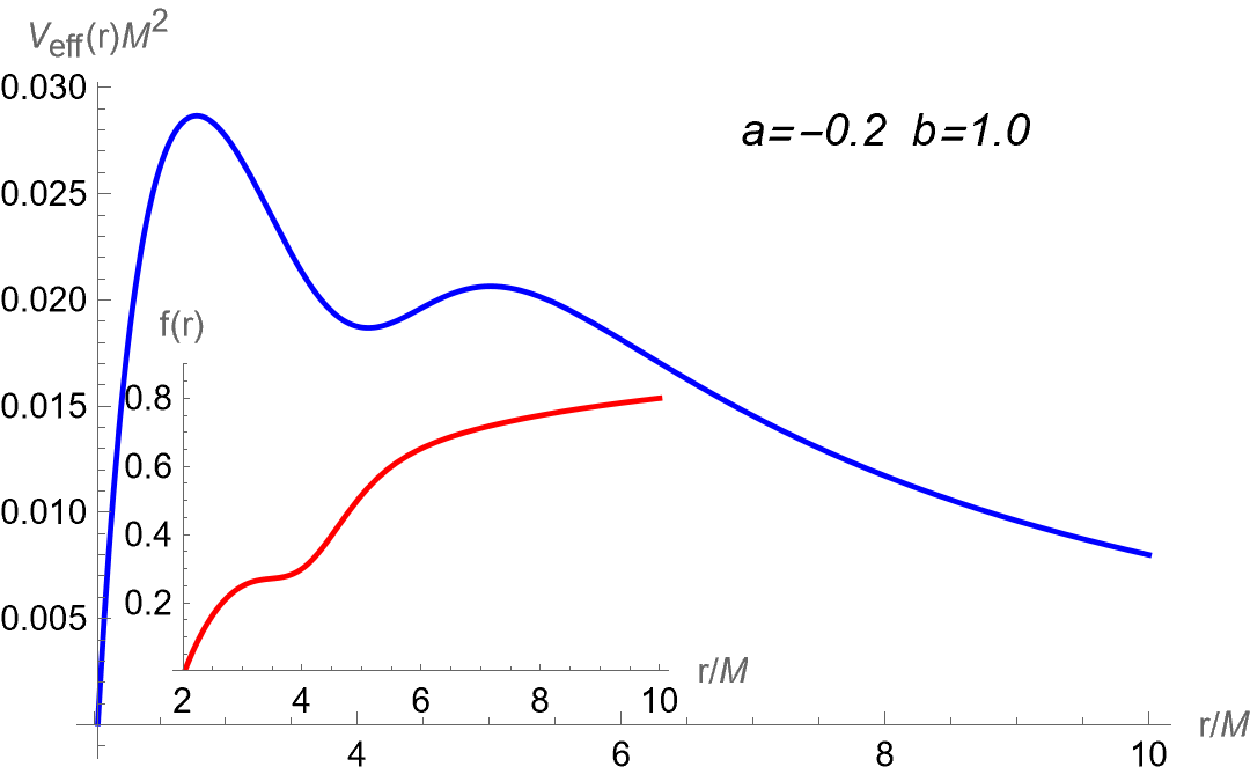}}\\
 \subfigure{\includegraphics[scale=0.48]{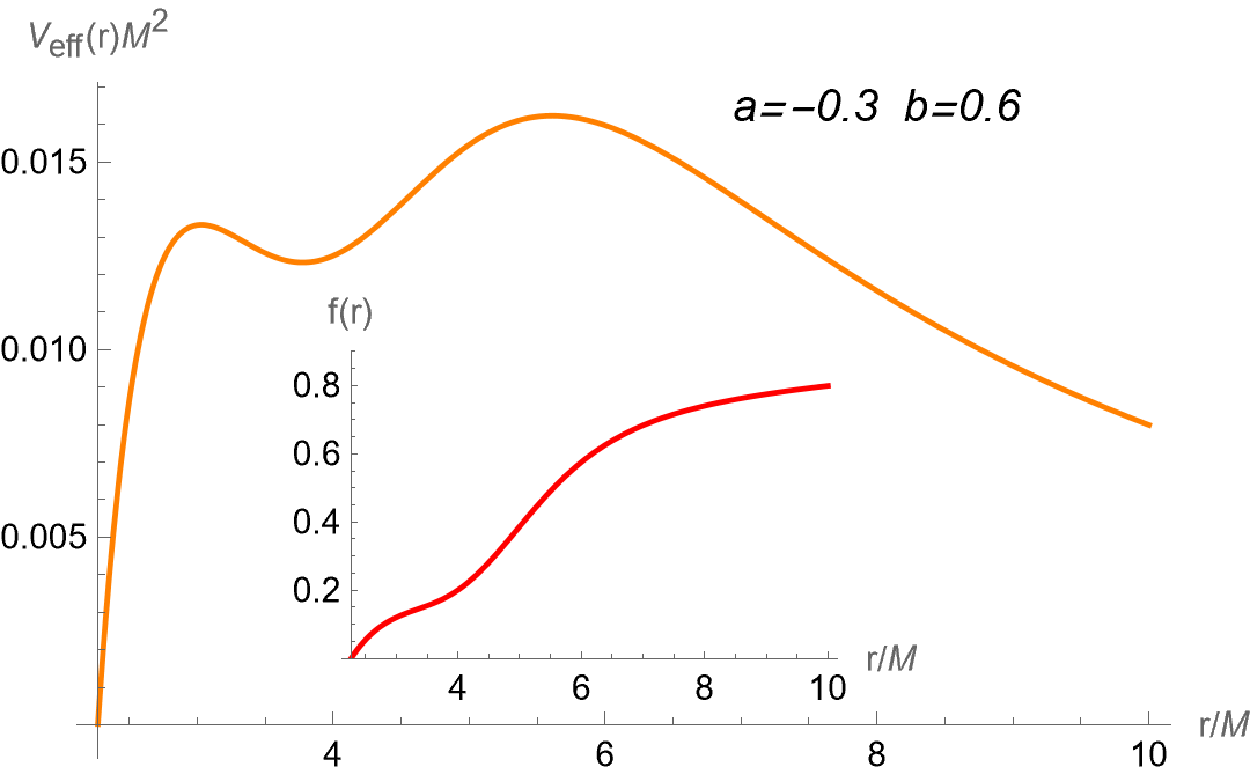}} %
\end{minipage}\caption{For the perturbed Schwarzschild metric $\left(\ref{eq:psbf}\right)$
with $r_{0}=4M$, there are two photon spheres outside the event horizon
in the blue and orange regions of the left panel. In the blue/orange
regions, the effective potential at the inner photon sphere is higher/lower
than that at the outer one. In the dark blue and orange regions, the
metric function $f\left(r\right)$ is monotonic, which is a necessary
condition for SEC. The right panels show that the effective potential
has a minimum, corresponding to an anti-photon sphere, near $r=4M$.}
\label{fig:PSB}
\end{figure}

Consider a small perturbation, which may be induced by environmental
effects, to the Schwarzschild metric. Specifically, we model the perturbation
by the Pöschl-Teller potential, and hence the perturbed Schwarzschild
metric has
\begin{equation}
f\left(r\right)=1-\frac{2M}{r}+\frac{a}{\cosh^{2}\left[b\left(r-r_{0}\right)/M\right]},\label{eq:psbf}
\end{equation}
where the parameters $a$, $b$ and $r_{0}$ describe the amplitude,
the width and the location of the perturbation, respectively. In the
left panel of FIG. \ref{fig:PSB}, colored regions, where two photon
spheres exist outside the event horizon, are displayed in the $a$-$b$
parameter space with $r_{0}=4M$:

\begin{itemize}
\item Blue regions: The potential peak at the inner photon sphere is higher
than that at the outer one. In this case, the two photon spheres can
both play a role in determining optical appearances of luminous matters
around black holes \cite{Gan:2021xdl,Gan:2021pwu,Guo:2022muy}.
\item Orange regions: The potential peak at the inner photon sphere is lower
than that at the outer one. In this case, light rays in the vicinity
of the inner photon sphere can not escape to infinity, making this
photon sphere invisible to distant observers. However, the inner photon
sphere is closely related to long-lived quasinormal modes \cite{Guo:2021enm}
and echo signals \cite{Guo:2022umh}.
\end{itemize}

Moreover, $f\left(r\right)$ is a monotonically increasing function
in the dark blue and orange regions of the left panel of FIG. \ref{fig:PSB}.
In addition, representative examples of the effective potential $V_{\text{eff}}(r)$
and the metric function $f\left(r\right)$ in the dark blue and orange
regions are given in the upper-right and lower-right panels of FIG.
\ref{fig:PSB}, respectively. It shows that $f\left(r\right)$ is
convex with $f^{\prime\prime}\left(r\right)>0$ around $r=4M$, where
$V_{\text{eff}}(r)$ has a minimum. If one assumes $f(r)=h(r)$, the
four energy conditions are found to be violated in the parameter space
of FIG. \ref{fig:PSB}.

\subsubsection{Johannsen-Psaltis Metric}

\begin{figure}[t]
\begin{centering}
\includegraphics[scale=0.6]{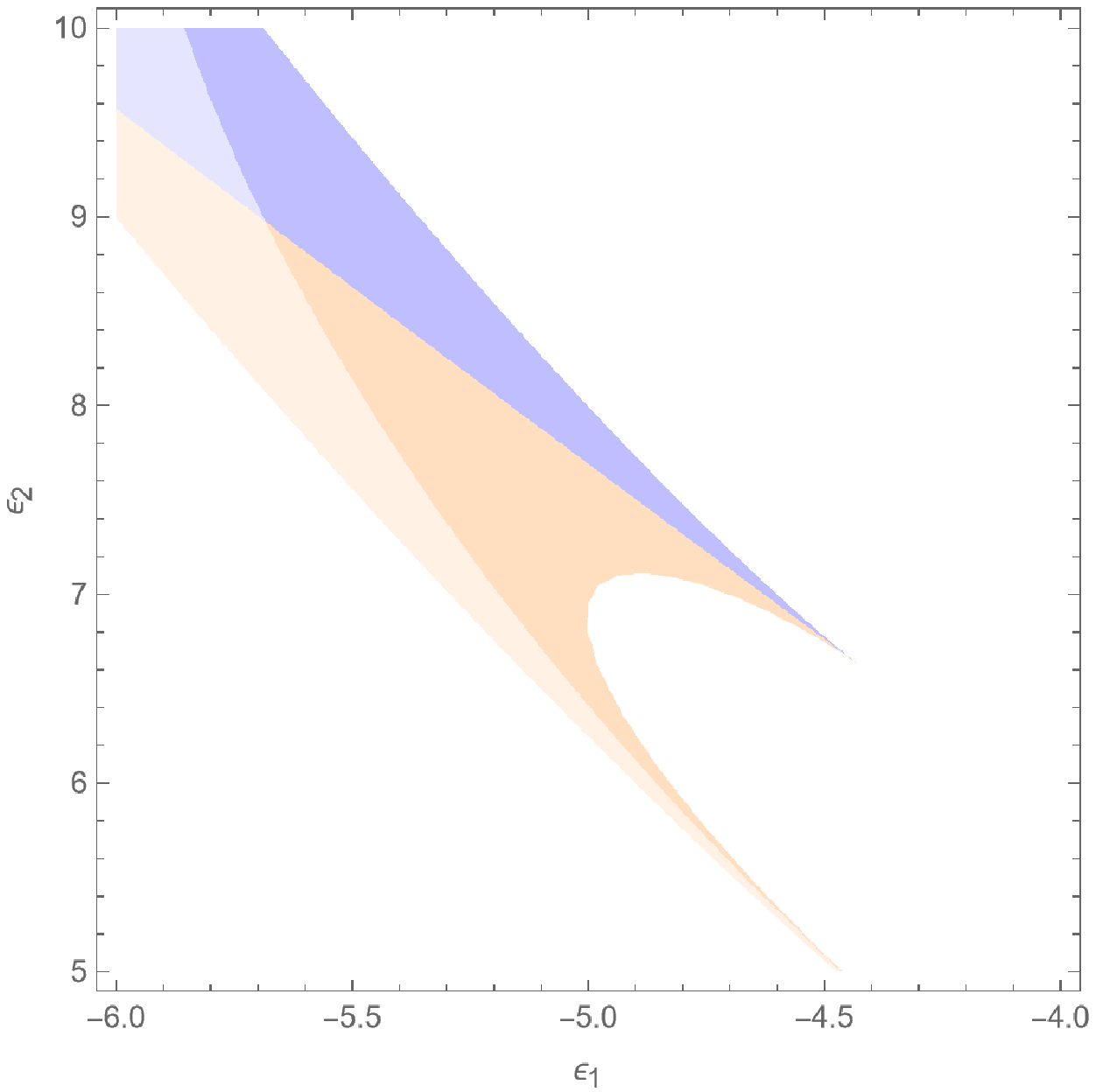} \includegraphics[scale=0.6]{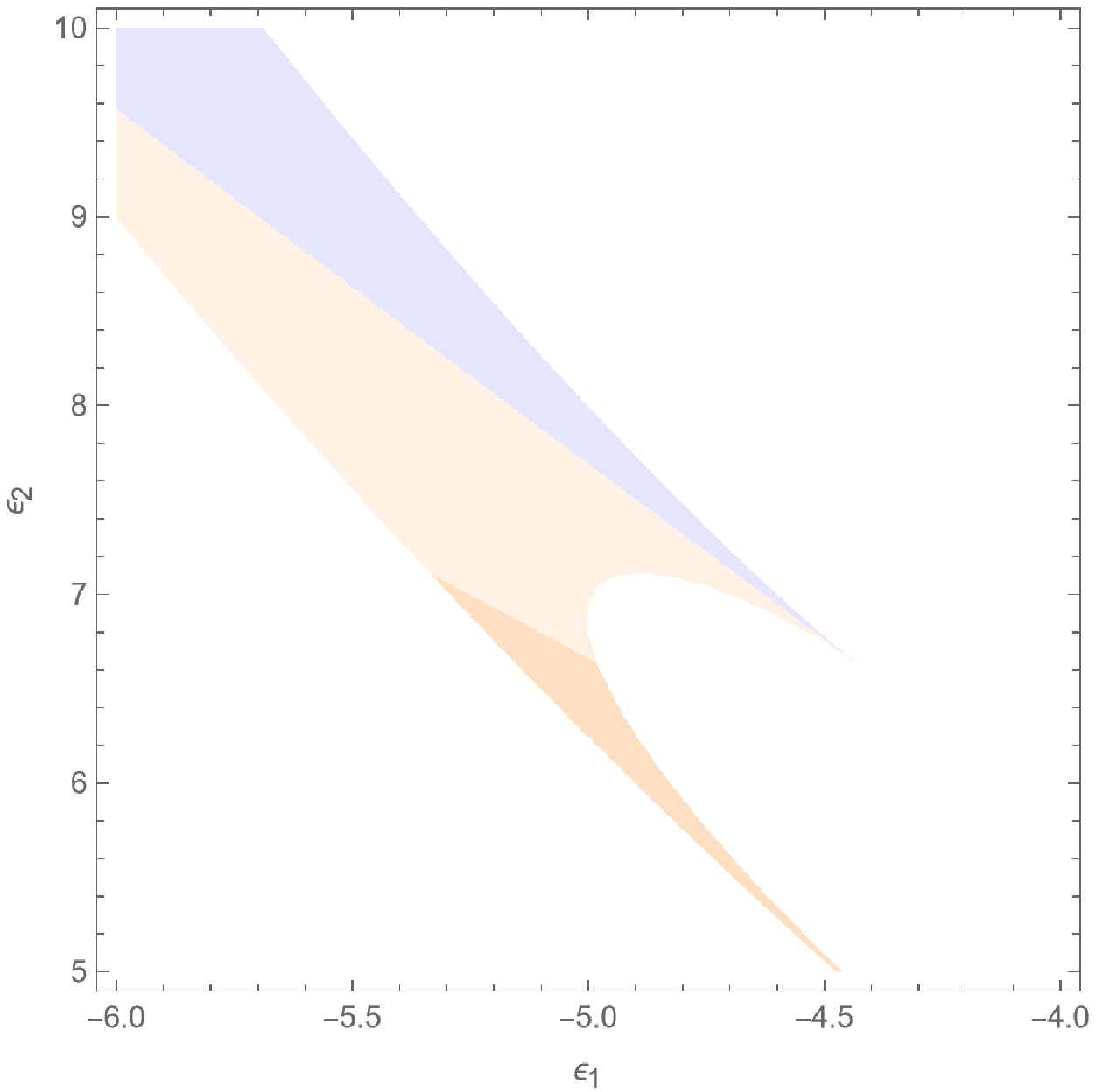}
\par\end{centering}
\caption{For the JP metric $\left(\ref{eq:JP}\right)$ with $\epsilon_{k>2}=0$,
two photon spheres exist outside the event horizon in the blue and
orange regions of the $\epsilon_{1}$-$\epsilon_{2}$ parameter space.
The inner potential peak is higher and lower than the outer one in
the blue and orange regions, respectively. In the dark blue and orange
regions, $f\left(r\right)$ is monotonic in the left panel, and NEC,
WEC and DEC are respected in the right panel, where $f(r)=h(r)$ is
assumed.}
\label{fig:JPEC}
\end{figure}

The metric function of the Johannsen-Psaltis (JP) metric can be written
as \cite{Johannsen:2011dh,Yunes:2011we}
\begin{equation}
f(r)=(1-\frac{2M}{r})\left[1+\sum_{k=1}^{\infty}\epsilon_{k}\left(\frac{M}{r}\right)^{k}\right],\label{eq:JP}
\end{equation}
where the coefficients $\epsilon_{k}$ are post-Newtonian-like parameters.
In FIG. \ref{fig:JPEC}, we focus on the case with $\epsilon_{k>2}=0$.
Blue and orange regions in the $\epsilon_{1}$-$\epsilon_{2}$ parameter
space denote parameter regions where two photon spheres exist. The
inner potential peak is higher/lower than the outer one in the blue/orange
regions. In the left panel, $f^{\prime}\left(r\right)>0$ in the dark
orange and blue regions. In the case with $f(r)=h(r)$, the JP metric
in the dark orange region of the right panel satisfies NEC, WEC and
DEC. However, SEC is always violated in the colored regions.

\subsubsection{New Parametrization}

\begin{figure}[t]
\begin{centering}
\includegraphics[scale=0.6]{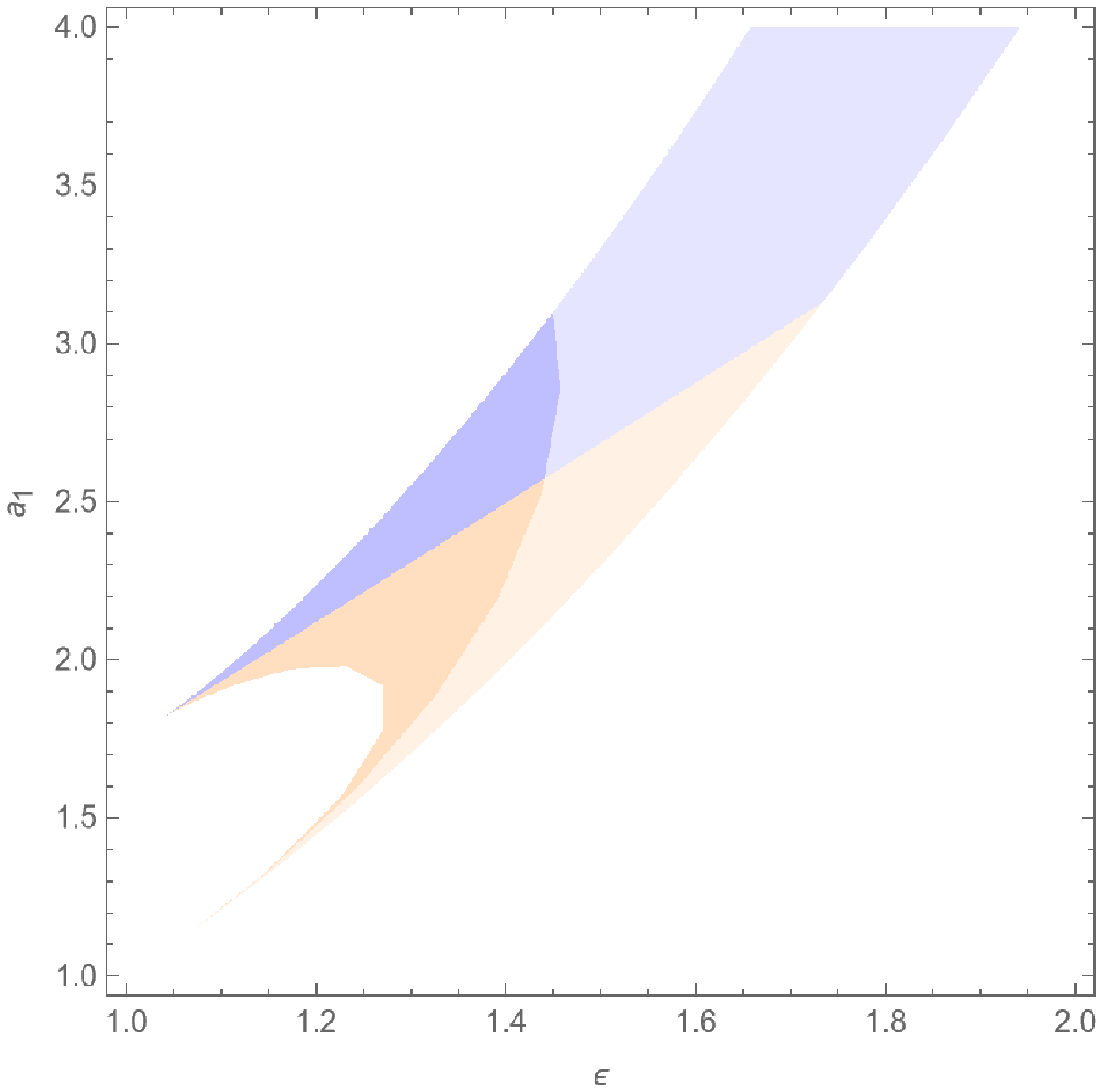} \includegraphics[scale=0.6]{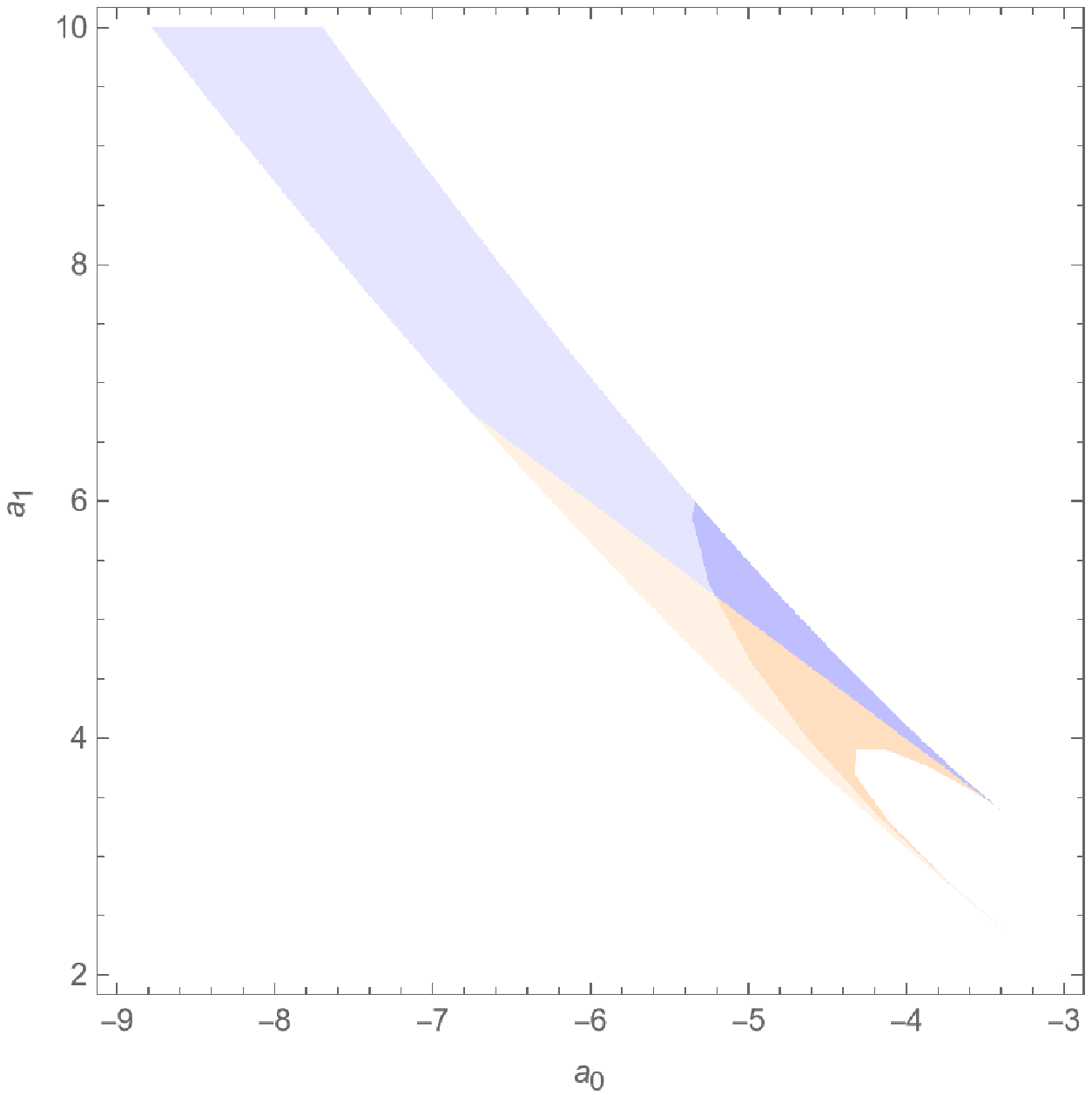}
\par\end{centering}
\caption{For the new parametrization $\left(\ref{eq:np}\right)$ with $a_{i>1}=0$,
two photon spheres exist in the blue and orange regions of the $\epsilon$-$a_{1}$
parameter space with $a_{0}=0$ (\textbf{Left}) and the $a_{0}$-$a_{1}$
parameter space with $\epsilon=0$ (\textbf{Right}). The inner potential
peak is higher and lower than the outer one in the blue and orange
regions, respectively. In the dark blue and orange regions, $f\left(r\right)$
is monotonically increasing.}
\label{fig:np}
\end{figure}

To better reproduce deviations from the Schwarzschild\ metric in
the strong-field regime, a new parametrization for spherically symmetric
black holes has been proposed \cite{Rezzolla:2014mua,Kocherlakota:2022jnz,Kocherlakota:2022mro},
\begin{equation}
f\left(r\right)=\left(1-\frac{r_{0}}{r}\right)\left[1-\epsilon\frac{r_{0}}{r}+\left(a_{0}-\epsilon\right)\frac{r_{0}^{2}}{r^{2}}+\tilde{A}\left(1-\frac{r_{0}}{r}\right)\frac{r_{0}^{3}}{r^{3}}\right],\label{eq:np}
\end{equation}
where $r_{0}$ is the horizon radius, and
\begin{equation}
\tilde{A}(x)=\frac{a_{1}}{1+\frac{a_{2}x}{1+\frac{a_{3}x}{1+\ldots}}}.
\end{equation}
The parameters $\epsilon$, $a_{0}$ and $a_{i}$ are used to characterize
the black hole metric. Current observational constraints on the PPN
parameters imply $a_{0}\sim10^{-4}$ \cite{Rezzolla:2014mua}. For
simplicity, we also set $a_{i>1}=0$. In FIG. \ref{fig:np}, we consider
two cases with $a_{0}=0$ and $\epsilon=0$ in the left and right
panels, respectively. Two photon spheres exist outside the event horizon
in the blue and orange regions, and the inner potential peak is higher
and lower than the outer one in the blue and orange regions, respectively.
In addition, $f\left(r\right)$ is monotonically increasing in the
dark blue and orange regions. The energy conditions are checked for
the colored regions assuming $f(r)=h(r)$, and it is found that only
NEC and WEC are respected in the small lower-left tip of the orange
region in the left panel.

\subsection{Specific Models}

\subsubsection{Einstein-Maxwell-Scalar Model}

Recently, a novel class of Einstein-Maxwell-scalar (EMS) models have
been proposed to study the spontaneous scalarization of RN black holes
\cite{Herdeiro:2018wub,Konoplya:2019goy,Wang:2020ohb,Guo:2021zed,Guo:2021ere}.
In such models, the scalar field can trigger a tachyonic instability
to form spontaneously scalarized hairy black holes from RN black holes.
The action of the EMS model is
\begin{equation}
S=\int d^{4}x\sqrt{-g}\left[\mathcal{R}-2\partial_{\mu}\phi\partial^{\mu}\phi-e^{\alpha\phi^{2}}F_{\mu\nu}F^{\mu\nu}\right],
\end{equation}
where $\mathcal{R}$ is the Ricci scalar, the scalar field $\phi$
is non-minimally coupled to the electromagnetic field $A_{\mu}$ with
the coupling function $e^{\alpha\phi^{2}}$, and $F_{\mu\nu}=\partial_{\mu}A_{\nu}-\partial_{\nu}A_{\mu}$
is the electromagnetic field strength tensor. Restricting to static
and spherically symmetric black hole solutions, one can have the generic
ansatz
\begin{align}
ds^{2} & =-N(r)e^{-2\delta(r)}dt^{2}+\frac{1}{N(r)}dr^{2}+r^{2}\left(d\theta^{2}+\sin^{2}\theta d\varphi^{2}\right),\nonumber \\
A_{\mu}dx^{\mu} & =V(r)dt\text{ and}\ \phi=\phi(r).\label{eq:HBH}
\end{align}
In the static orthonormal basis, we find
\begin{align}
T_{00} & =2N\left(r\right)\left[\phi^{\prime}\left(r\right)\right]^{2}+2e^{2\delta\left(r\right)+\alpha\phi^{2}\left(r\right)}\left[V^{\prime}\left(r\right)\right]^{2}=\rho,\nonumber \\
T_{11} & =2N\left(r\right)\left[\phi^{\prime}\left(r\right)\right]^{2}-2e^{2\delta\left(r\right)+\alpha\phi^{2}\left(r\right)}\left[V^{\prime}\left(r\right)\right]^{2}=p_{1},\nonumber \\
T_{22} & =-2N\left(r\right)\left[\phi^{\prime}\left(r\right)\right]^{2}+2e^{2\delta\left(r\right)+\alpha\phi^{2}\left(r\right)}\left[V^{\prime}\left(r\right)\right]^{2}=p_{2},\\
T_{33} & =-2N\left(r\right)\left[\phi^{\prime}\left(r\right)\right]^{2}+2e^{2\delta\left(r\right)+\alpha\phi^{2}\left(r\right)}\left[V^{\prime}\left(r\right)\right]^{2}=p_{3}.\nonumber
\end{align}
Using the aforementioned criteria for the energy conditions, it is
easily found that NEC, WEC, DEC and SEC are all satisfied for the
black hole solution $\left(\ref{eq:HBH}\right)$ in the EMS model.

To obtain the black hole solution $\left(\ref{eq:HBH}\right)$, one
needs to impose proper boundary conditions at the event horizon $r_{h}$
and spatial infinity,
\begin{align}
N(r_{h}) & =0\text{, }\delta(r_{h})=\delta_{0}\text{, }\phi(r_{h})=\phi_{0}\text{, }V(r_{h})=0\text{,}\nonumber \\
N(\infty) & =1\text{,}\ \delta(\infty)=0\text{, }\phi(\infty)=0\text{, }V(\infty)=\Phi\text{,}\label{eq:boundary conditions}
\end{align}
where $\delta_{0}$ and $\phi_{0}$ can be used to characterize black
hole solutions, and $\Phi$ is the electrostatic potential. Specifically,
$\phi_{0}=\delta_{0}=0$ correspond to the scalar-free solutions with
$\phi=0$, i.e., RN black holes. When non-zero values of $\phi_{0}$
and $\delta_{0}$ are admitted, scalarized RN black holes with a non-trivial
scalar field $\phi$ can be obtained by a shooting method.

\begin{figure}[t]
\centering%
\begin{minipage}[c]{0.45\linewidth}%
 \subfigure{\includegraphics[scale=0.45]{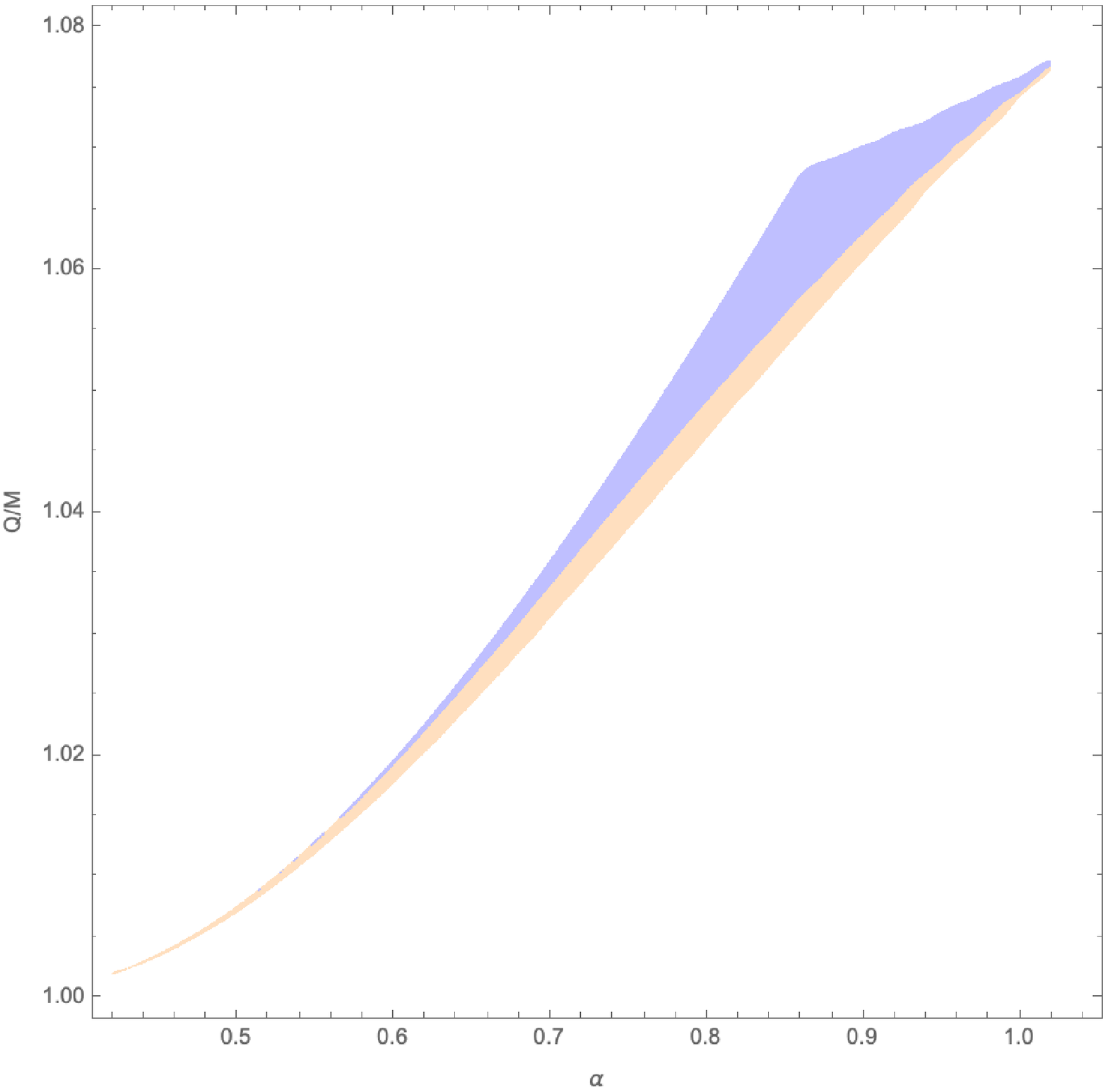}} %
\end{minipage}\hspace{2pt} %
\begin{minipage}[c]{0.45\linewidth}%
 \subfigure{\includegraphics[scale=0.64]{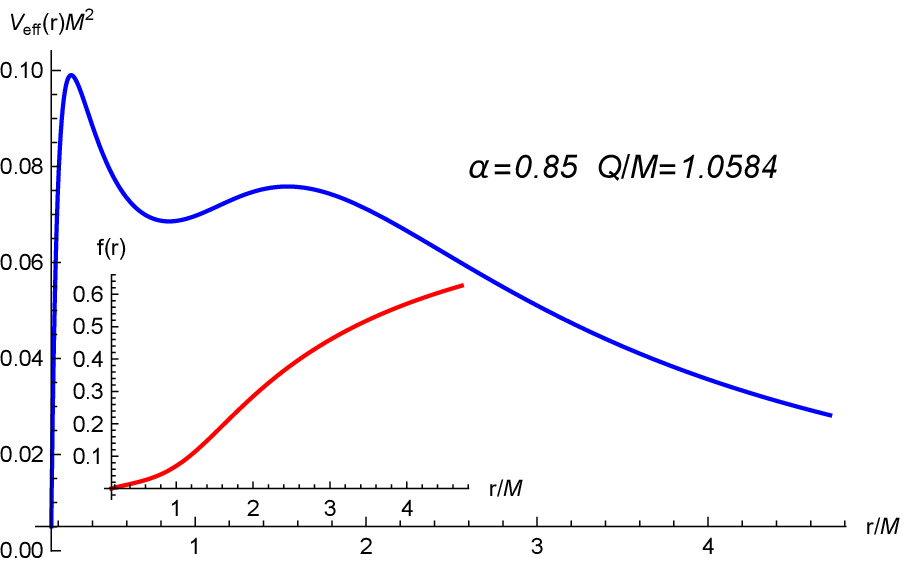}}\\
 \subfigure{\includegraphics[scale=0.64]{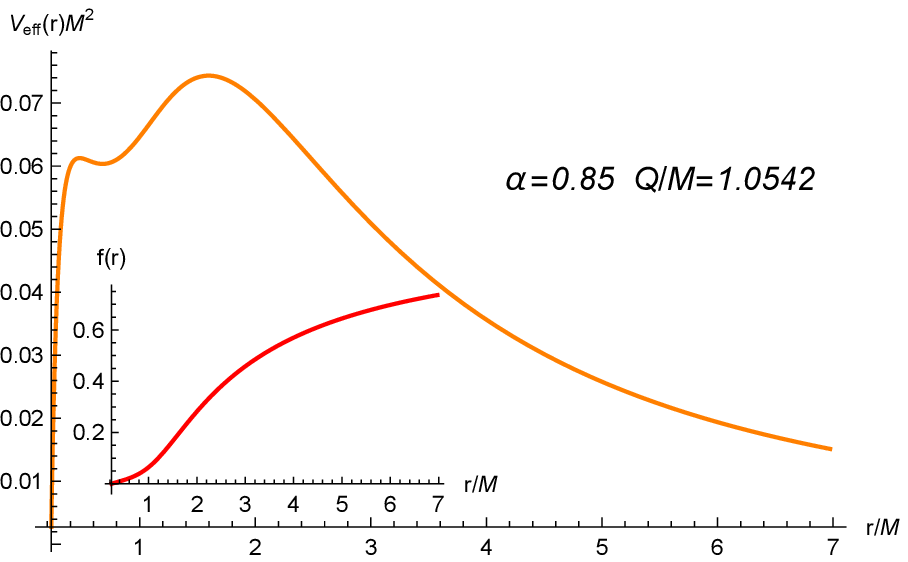}} %
\end{minipage}\caption{\textbf{Left}:\ The parameter regions of scalarized RN black hole
$\left(\ref{eq:HS}\right)$ that double photon spheres exist. In the
blue/orange region, the effective potential at the inner photon sphere
is higher/lower than that at the outer one. As discussed in the text,
NEC, WEC, DEC and SEC are always satisfied. \textbf{Right}: The metric
function $f\left(r\right)=N(r)e^{-2\delta(r)}$ and the effective
potential $V_{eff}\left(r\right)$ of representative black holes in
the blue and orange regions are presented in the upper and lower panels,
respectively. It shows that $f\left(r\right)$ is monotonically increasing,
which is a necessary condition for the validity of SEC.}
\label{fig:HBH}
\end{figure}

In the left panel of FIG. \ref{fig:HBH}, we display blue and orange
regions in the $\alpha$-$Q/M$ parameter space, in which double photon
spheres exist outside the event horizon. Here, the parameters $M$
and $Q$ are the black hole mass and charge, respectively. In the
blue/orange region, the potential peak at the inner/outer photon sphere
is higher than that at the outer/inner photon sphere. The right panels
of FIG. \ref{fig:HBH} present the metric function $f\left(r\right)=N(r)e^{-2\delta(r)}$
and the effective potential $V_{\text{eff}}\left(r\right)$ for black
holes in the blue and orange regions. As expected, $f\left(r\right)$
is shown to monotonically increase, which is required by the validity
of SEC.

\subsubsection{Black Holes in Galaxies}

\begin{figure}[ptbh]
\begin{centering}
\includegraphics[scale=0.6]{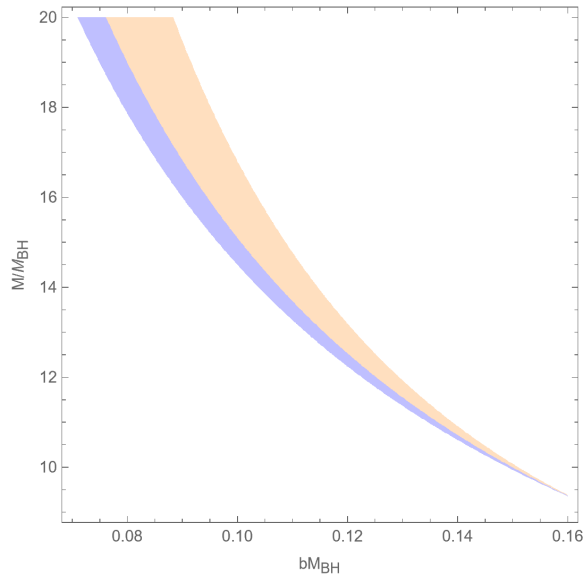}
\par\end{centering}
\caption{For the metric describing black holes in galaxies $\left(\ref{eq:bhG}\right)$,
two photon spheres exist in the blue and orange regions of the $bM_{\text{BH}}$-$M/M_{\text{BH}}$
parameter space. The inner potential peak is higher than the outer
one in the blue region, while the outer potential peak is higher than
the inner one in the orange region.}
\label{fig:BG}
\end{figure}

To describe the geometry of supermassive black holes residing in galaxies,
a family of solutions of Einstein's gravity has been constructed by
minimally coupling the gravity sector to an anisotropic fluid \cite{Cardoso:2021wlq}.
In particular, the geometry can be written as
\begin{align}
ds^{2} & =-\left(1-\frac{2M_{BH}}{r}\right)e^{\Upsilon}dt^{2}+\frac{dr^{2}}{1-2m(r)/r}+r^{2}d\Omega^{2},\nonumber \\
\Upsilon & =-\pi\sqrt{\frac{M}{\xi}}+2\sqrt{\frac{M}{\xi}}\arctan\frac{r+b^{-1}-M}{\sqrt{M\xi}},\nonumber \\
\xi & =2b^{-1}-M+4M_{\mathrm{BH}},\label{eq:bhG}\\
m\left(r\right) & =M_{\mathrm{BH}}+\frac{Mr^{2}}{\left(b^{-1}+r\right)^{2}}\left(1-\frac{2M_{\mathrm{BH}}}{r}\right)^{2},\nonumber
\end{align}
where $M_{\mathrm{BH}}$ is the black hole mass, and $M$ and $b^{-1}$
are the total mass and the typical lengthscale of the halo surrounding
the black hole. When $M\rightarrow0$ and $b\rightarrow0$, the above
metric reduces to the Schwarzschild metric. Moreover, it showed that
the anisotropic fluid satisfies NEC, WEC and SEC, and DEC is only
violated close to the event horizon. Nevertheless, the anisotropic
fluid has arbitrarily small pressure and density near the horizon,
and hence the violation of DEC may play no role in the spacetime dynamics
\cite{Cardoso:2021wlq}. In FIG. \ref{fig:BG}, we present blue and
orange regions in the $bM_{\text{BH}}$-$M/M_{\text{BH}}$ parameter
space, which admit the black hole solutions with double photon spheres.
In the blue/orange region, the potential peak at the inner/outer photon
sphere is higher than that at the outer/inner photon sphere.

\subsubsection{Hairy Schwarzschild Black Holes}

\begin{figure}[t]
\begin{centering}
\includegraphics[scale=0.6]{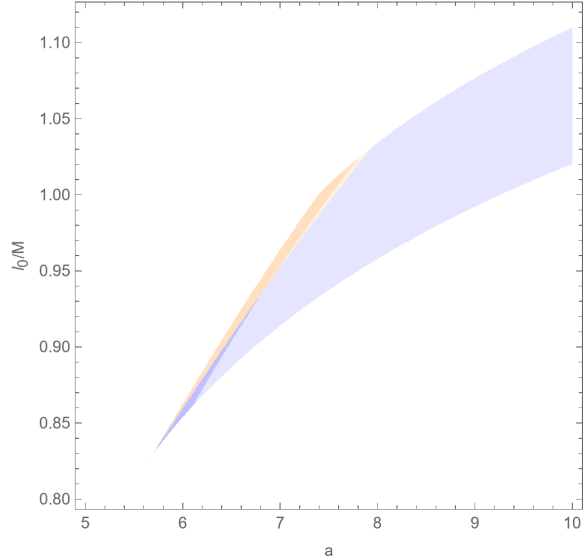}
\par\end{centering}
\caption{The parameter region of hairy Schwarzschild black holes $\left(\ref{eq:HS}\right)$
that double photon spheres exist. NEC and WEC are respected in the
dark blue and orange regions. The inner potential peak is higher and
lower than the outer one in the blue and orange regions, respectively.}
\label{fig:HS}
\end{figure}

In \cite{Ovalle:2017fgl,Ovalle:2018gic,Ovalle:2020kpd}, a hairy black
hole solution was obtained from the seed Schwarzschild vacuum by the
gravitational decoupling approach. The metric of the hairy black hole
is
\begin{equation}
ds^{2}=-\left[1-\frac{2M}{r}+ae^{-r/\left(M-l_{0}/2\right)}\right]dt^{2}+\frac{dr^{2}}{1-\frac{2M}{r}+ae^{-r/\left(M-l_{0}/2\right)}}+r^{2}\left(d\theta^{2}+\sin^{2}\theta d\varphi^{2}\right),\label{eq:HS}
\end{equation}
where $M$ is the black hole mass, $a$ is the deviation parameter,
and $l_{0}<2M$. When $a\rightarrow0$, this metric reduces to the
Schwarzschild metric. Moreover, the effective density and radial pressure
are
\begin{equation}
\rho=-p_{1}=\frac{ae^{-r/\left(M-l_{0}/2\right)}}{\left(M-l_{0}/2\right)r^{2}}\left(r-M+l_{0}/2\right),\label{eq:HSrho}
\end{equation}
and the effective tangential pressures are
\begin{equation}
p_{2}=p_{3}=\frac{ae^{-r/\left(M-l_{0}/2\right)}}{2\left(M-l_{0}/2\right)^{2}r}\left(r-2M+l_{0}\right).\label{eq:HSp}
\end{equation}

In the left panel of FIG. \ref{fig:HS}, we display colored regions
in the $a$-$l_{0}/M$ parameter space, which admit the black hole
solutions with double photon spheres. In the blue/orange regions,
the potential peak at the inner/outer photon sphere is higher than
that at the outer/inner photon sphere. Moreover, we also check whether
the black hole solutions satisfy the energy conditions. It is found
that DEC and SEC are always violated. Nevertheless, NEC and WEC can
be respected in dark blue and orange regions in FIG. \ref{fig:HS}.
From eqns. $\left(\ref{eq:HSrho}\right)$ and $\left(\ref{eq:HSp}\right)$,
the violations of NEC, WEC and SEC occur near the event horizon.

\section{NLED Black Holes}

\label{sec:NLEDBH}

If quantum corrections are considered, non-linear terms are usually
added to the Maxwell Lagrangian, which gives rise to an effective
model, namely non-linear electrodynamics (NLED). In the Einstein-NLED
theory, various charged black hole solutions were investigated \cite{Soleng:1995kn,AyonBeato:1998ub,Maeda:2008ha,Fan:2016hvf,Hendi:2017mgb,Tao:2017fsy,Wang:2019jzz}.
Due to self-interactions, photons usually do not propagate along null
geodesics of the underlying black hole spacetime. Instead, they move
along null geodesics of some effective geometry \cite{Novello:1999pg}.
Therefore, we investigate the existence of double photon spheres in
the effective metric of NLED black holes in this section.

Consider a $\left(3+1\right)$ dimensional model of gravity coupled
to a non-linear electromagnetic field $A_{\mu}$,
\begin{equation}
\mathcal{S}=\int d^{4}x\sqrt{-g}\left[R+4\mathcal{L}\left(s,p\right)\right],\label{eq:NLEDAction}
\end{equation}
where $\mathcal{L}\left(s,p\right)$ is a general NLED Lagrangian.
Here, $s$ and $p$ are two independent nontrivial scalars built from
the field strength tensor $F_{\mu\nu}=\partial_{\mu}A_{\nu}-\partial_{\nu}A_{\mu}$
and none of its derivatives,
\begin{equation}
s=-\frac{1}{4}F^{\mu\nu}F_{\mu\nu}\text{ and }p=-\frac{1}{8}\epsilon^{\mu\nu\rho\sigma}F_{\mu\nu}F_{\rho\sigma}\text{,}
\end{equation}
where $\epsilon^{\mu\nu\rho\sigma}\equiv-\left[\mu\text{ }\nu\text{ }\rho\text{ }\sigma\right]/\sqrt{-g}$
is a totally antisymmetric Lorentz tensor, and $\left[\mu\text{ }\nu\text{ }\rho\text{ }\sigma\right]$
is the permutation symbol. Varying the action $\left(\ref{eq:NLEDAction}\right)$
in terms of $g_{\mu\nu}$ and $A_{\mu}$ leads to equations of motion,
\begin{align}
R_{\mu\nu}-\frac{1}{2}Rg_{\mu\nu} & =\frac{T_{\mu\nu}}{2}\text{,}\nonumber \\
\nabla_{\mu}\left[\frac{\partial\mathcal{L}\left(s,p\right)}{\partial s}F^{\mu\nu}+\frac{1}{2}\frac{\partial\mathcal{L}\left(s,p\right)}{\partial p}\epsilon^{\mu\nu\rho\sigma}F_{\rho\sigma}\right] & =0\text{,}\label{eq:NLEDEOM}
\end{align}
where $T_{\mu\nu}$ is the energy-momentum tensor for the NLED field,
\begin{equation}
T_{\mu\nu}=4g_{\mu\nu}\left[\mathcal{L}\left(s,p\right)-p\frac{\partial\mathcal{L}\left(s,p\right)}{\partial p}\right]+\frac{\partial\mathcal{L}\left(s,p\right)}{\partial s}F_{\mu}^{\text{ }\rho}F_{\nu\rho}\text{.}
\end{equation}

Here, we focus on the spherically symmetric and static black hole
solution,
\begin{align}
ds^{2} & =-f\left(r\right)dt^{2}+\frac{dr^{2}}{f\left(r\right)}+r^{2}\left(d\theta^{2}+\sin^{2}\theta d\varphi^{2}\right)\text{,}\nonumber \\
A & =A_{t}\left(r\right)dt-P\cos\theta d\varphi,\label{eq:NLEDBH}
\end{align}
where $P$ is the black hole magnetic charge. Plugging the ansatz
$\left(\ref{eq:NLEDBH}\right)$ into eqn. $\left(\ref{eq:NLEDEOM}\right)$
yields
\begin{equation}
f\left(r\right)=1-\frac{2M}{r}-\frac{2}{r}\int_{r}^{\infty}r^{2}\left[\mathcal{L}\left(s,p\right)-A_{t}^{\prime}\left(r\right)\frac{Q}{r^{2}}\right]dr,\label{eq:NLEDf}
\end{equation}
where $A_{t}^{\prime}\left(r\right)$ is determined by
\begin{equation}
\frac{\partial\mathcal{L}\left(s,p\right)}{\partial s}A_{t}^{\prime}\left(r\right)-\frac{\partial\mathcal{L}\left(s,p\right)}{\partial p}\frac{P}{r^{2}}=\frac{Q}{r^{2}}\text{ with }s=\frac{A_{t}^{\prime2}\left(r\right)-P^{2}/r^{4}}{2}\text{ and }p=-\frac{A_{t}^{\prime}\left(r\right)P}{r^{2}}.
\end{equation}
Here, $M$ and $Q$ are the black hole mass and electrical charge,
respectively. In the static orthonormal basis, the energy density
and the pressures are
\begin{align}
\rho & =-p_{1}=\frac{2QA_{t}^{\prime}\left(r\right)}{r^{2}}-2\mathcal{L}\left(s,p\right),\nonumber \\
p_{2} & =p_{3}=2\mathcal{L}\left(s,p\right)+\frac{2A_{t}^{\prime}\left(r\right)P}{r^{2}}\frac{\partial\mathcal{L}\left(s,p\right)}{\partial p}+\frac{2P^{2}}{r^{4}}\frac{\partial\mathcal{L}\left(s,p\right)}{\partial s}.
\end{align}

In \cite{Novello:1999pg}, characteristic equations of NLED fields
were used to derive the effective geometry of photons in the underlying
NLED black hole spacetime. If the NLED Lagrangian $\mathcal{L}_{\text{NLED}}$
is a function of $s$, i.e., $\mathcal{L}_{\text{NLED}}\mathcal{=L}\left(s\right)$,
the effective metric is
\begin{equation}
g_{\text{eff}}^{\mu\nu}=-4\frac{\partial\mathcal{L}\left(s\right)}{\partial s}g^{\mu\nu}-64\frac{\partial^{2}\mathcal{L}\left(s\right)}{\partial s^{2}}F_{\text{ }\alpha}^{\mu}F^{\alpha\nu}.
\end{equation}
For $\mathcal{L}_{\text{NLED}}\mathcal{=L}\left(s,p\right)$, the
full expression of $g_{\text{eff}}^{\mu\nu}$ is too lengthy to be
included here. We refer readers to \cite{Novello:1999pg} for the
expression of $g_{\text{eff}}^{\mu\nu}$. In the NLED black hole $\left(\ref{eq:NLEDBH}\right)$,
the effective metric is spherically symmetric,
\begin{equation}
ds^{2}=g_{\text{eff}}^{\mu\nu}dx_{\mu}dx_{\nu}=-f_{\text{eff}}\left(r\right)dt^{2}+\frac{dr^{2}}{h_{\text{eff}}\left(r\right)}+R_{\text{eff}}\left(r\right)\left(d\theta^{2}+\sin^{2}\theta d\varphi^{2}\right),
\end{equation}
which gives the effective potential of photons in the effective metric,
\begin{equation}
V_{\text{eff}}\left(r\right)=\frac{f_{\text{eff}}\left(r\right)}{R_{\text{eff}}\left(r\right)}.
\end{equation}

\subsection{Quasi-topological Electromagnetism}

In \cite{Liu:2019rib}, the squared norm of the topological 4-form
$F\wedge F$ is added to the Maxwell theory, leading to the Quasi-topological
Electromagnetism (QE). In particular, the electrodynamics Lagrangian
is
\begin{equation}
\mathcal{L}\left(s,p\right)=s-aU^{\left(2\right)}=s+8ap^{2},
\end{equation}
where $U^{\left(2\right)}=-\epsilon^{\nu_{1}\nu_{2}\nu_{3}\nu_{4}}\epsilon_{\mu_{1}\mu_{2}\mu_{3}\mu_{4}}F^{\mu_{1}\mu_{2}}F^{\mu_{3}\mu_{4}}F_{\nu_{1}\nu_{2}}F_{\nu_{3}\nu_{4}}/8$
is the squared norm of the topological structure $F\wedge F$. Using
eqn. $\left(\ref{eq:NLEDf}\right)$, one can obtain the the metric
function $f\left(r\right)$,
\begin{equation}
f\left(r\right)=1-\frac{2M}{r}+\frac{P^{2}}{r^{2}}+\frac{Q^{2}}{r^{2}}\text{ }_{2}F_{1}\left(\frac{1}{4},1,\frac{5}{4};-\frac{16aP^{2}}{r^{4}}\right),\label{eq:QEBH}
\end{equation}
where $_{2}F_{1}\left(a,b,c;x\right)$ is the hypergeometric function.
Moreover, the energy density and the pressures are
\begin{align}
\rho & =-p_{1}=\frac{2P^{2}}{r^{4}}+\frac{2Q^{2}}{16aP^{2}+r^{4}},\nonumber \\
p_{2} & =p_{3}=\rho-\frac{64aP^{2}Q^{2}}{\left(16aP^{2}+r^{4}\right)^{2}},
\end{align}
which shows that NEC, WEC and DEC are always satisfied. Nevertheless,
SEC is not necessarily satisfied.

\begin{figure}[t]
\begin{centering}
\includegraphics[scale=0.6]{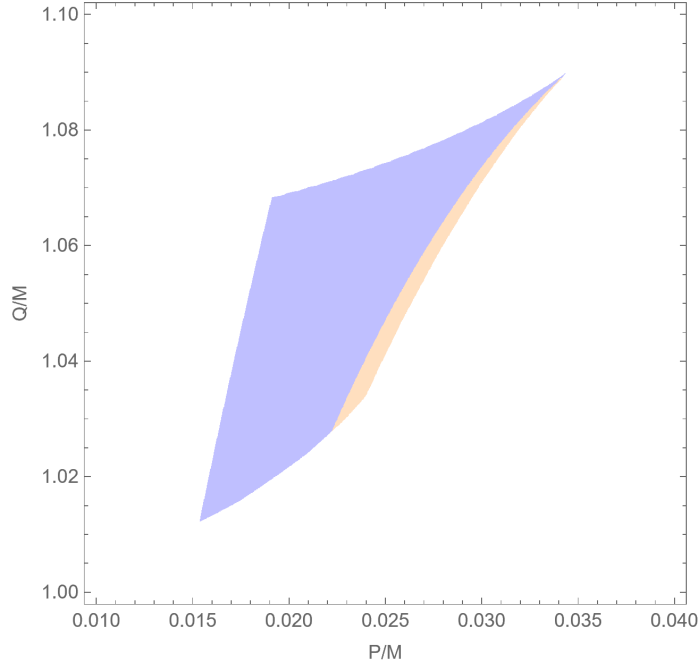}
\par\end{centering}
\caption{For the dyonic black holes with a quasi-topological electromagnetic
term $\left(\ref{eq:QEBH}\right)$, two photon spheres exist in the
blue and orange regions of the $P/M$-$Q/M$ parameter space with
$a=123.892$. The inner potential peak is higher and lower than the
outer one in the blue and orange regions, respectively.}
\label{fig:QE}
\end{figure}

Since $\partial^{2}\mathcal{L}\left(s,p\right)/\partial s^{2}=0=\partial^{2}\mathcal{L}\left(s,p\right)/\partial s\partial p$
in the QE, the effective metric of photons is just the underlying
spacetime $\left(\ref{eq:QEBH}\right)$. In other words, the non-linear
term in the quasi-topological electromagnetism does not induce a modification
on the spacetime metric for photons. In FIG. \ref{fig:QE}, we display
colored regions, where the QE black holes have two photon spheres.
Moreover, we find that SEC is always violated even though $f\left(r\right)$
can monotonically increase in some parameter region.

\subsection{Born-Infeld Electrodynamics}

The Born-Infeld (BI) electrodynamics was first proposed to smooth
divergences of the electrostatic self-energy of point charges by introducing
a cutoff on electric fields \cite{Born:1934gh}. Later, it is realized
that BI electrodynamics can emerge from the low energy limit of string
theory, which encodes the low-energy dynamics of D-branes. Coupling
the BI electrodynamics field to gravity, the BI black hole solution
was first obtained in \cite{Dey:2004yt,Cai:2004eh}. The BI electrodynamics
is described by the Lagrangian density
\begin{equation}
\mathcal{L}\left(s,p\right)=\frac{1}{a}\left(1-\sqrt{1-2as-a^{2}p^{2}}\right)\text{,}\label{eq:BI}
\end{equation}
where the coupling parameter $a$ is related to the string tension
$\alpha^{\prime}$ as $a=\left(2\pi\alpha^{\prime}\right)^{2}$. Eqn.
$\left(\ref{eq:NLEDf}\right)$ gives
\begin{equation}
f\left(r\right)=1-\frac{2M}{r}-\frac{2\left(P^{2}+Q^{2}\right)}{3\sqrt{r^{4}+a\left(P^{2}+Q^{2}\right)}+3r^{2}}+\frac{4\left(P^{2}+Q^{2}\right)}{3r^{2}}\text{ }_{2}F_{1}\left(\frac{1}{4},\frac{1}{2},\frac{5}{4};-\frac{a\left(P^{2}+Q^{2}\right)}{r^{4}}\right).\label{eq:BIBHf(r)}
\end{equation}
The energy density and the pressures are
\begin{align}
\rho & =-p_{1}=\frac{4\left(P^{2}+Q^{2}\right)}{r^{2}\left[\sqrt{a\left(P^{2}+Q^{2}\right)+r^{4}}+r^{2}\right]},\nonumber \\
p_{2} & =p_{3}=\frac{4\left(P^{2}+Q^{2}\right)}{\sqrt{a\left(P^{2}+Q^{2}\right)+r^{4}}\left(\sqrt{a\left(P^{2}+Q^{2}\right)+r^{4}}+r^{2}\right)},
\end{align}
which indicate that NEC, WEC, DEC and SEC are always satisfied. Furthermore,
the effective metric of photons is \cite{He:2022opa}
\begin{align}
f_{\text{eff}}\left(r\right) & =h_{\text{eff}}^{-1}\left(r\right)=\frac{\sqrt{aP^{2}+r^{4}}}{r^{2}}\left[\frac{aP^{2}+r^{4}}{a\left(P^{2}+Q^{2}\right)+r^{4}}\right]^{3/2}f\left(r\right),\nonumber \\
R_{\text{eff}}\left(r\right) & =\frac{\left(aP^{2}+r^{4}\right)^{2}}{r^{4}\sqrt{a\left(P^{2}+Q^{2}\right)+r^{4}}}.\label{eq:BIeffM}
\end{align}

\begin{figure}[t]
\centering%
\begin{minipage}[c]{0.45\linewidth}%
 \subfigure{\includegraphics[scale=0.56]{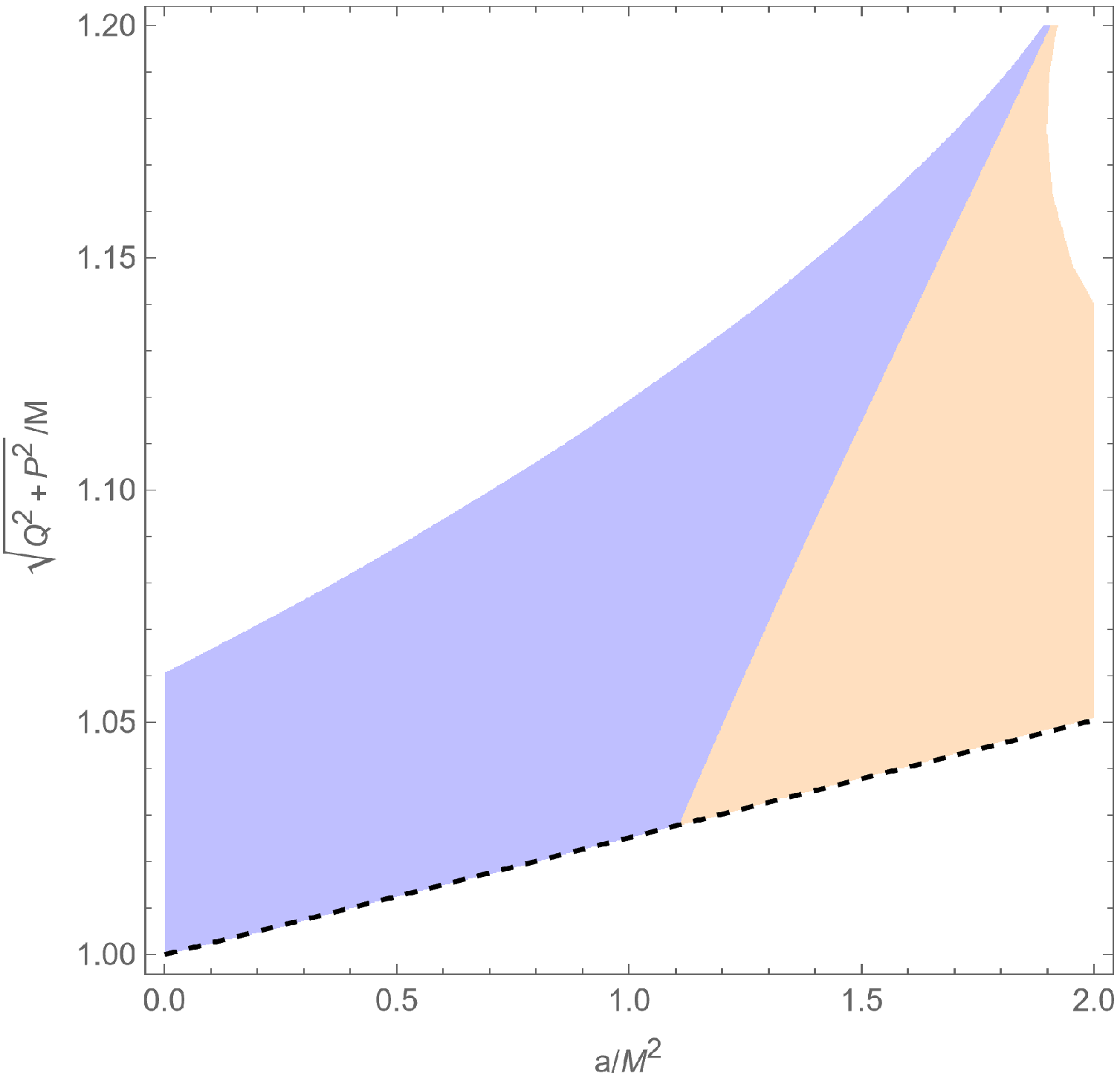}} %
\end{minipage}\hspace{20pt} %
\begin{minipage}[c]{0.45\linewidth}%
 \subfigure{\includegraphics[scale=0.48]{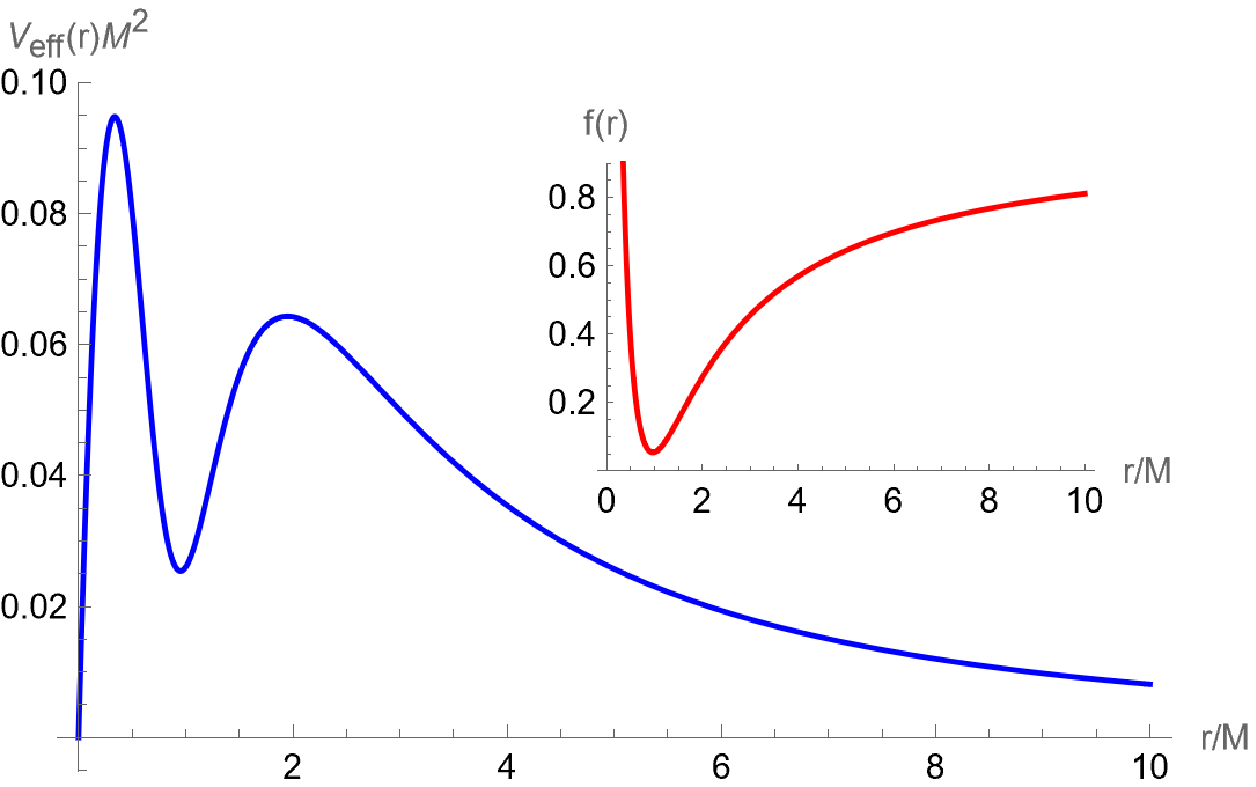}}\\
 \subfigure{\includegraphics[scale=0.48]{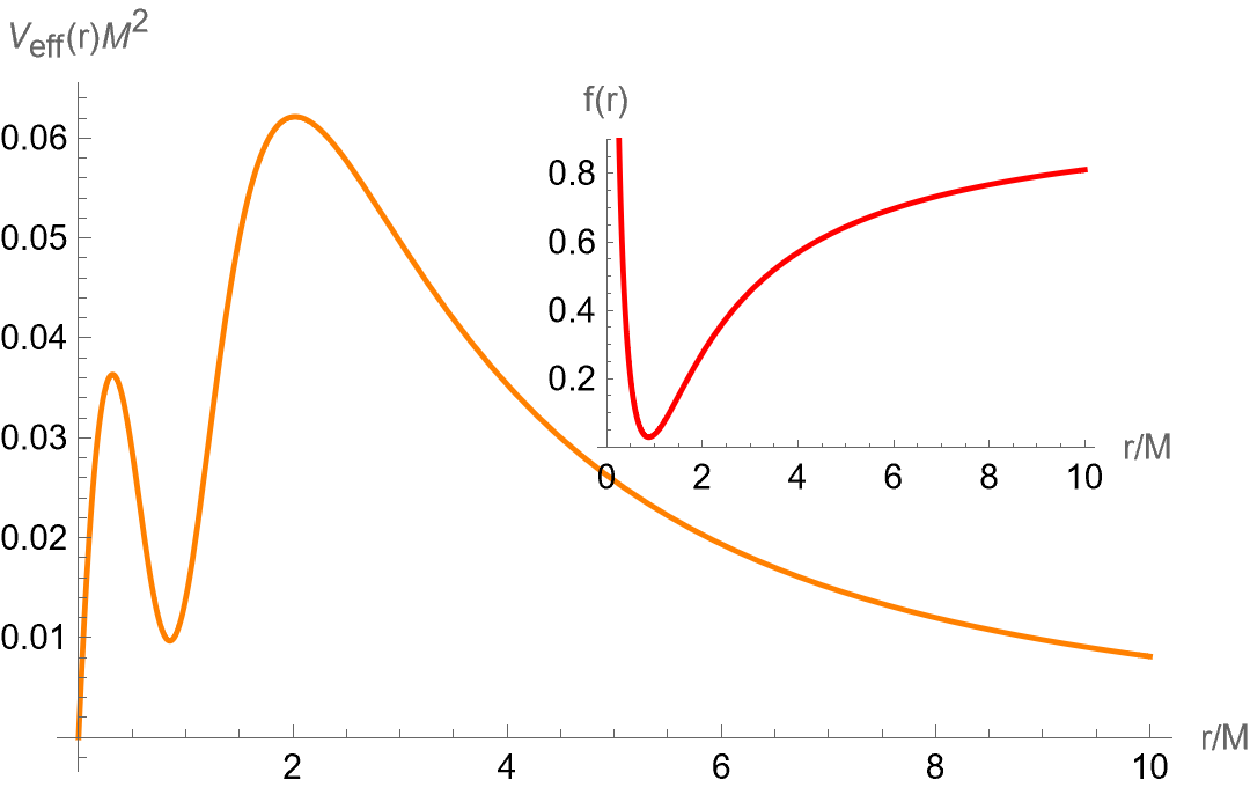}} %
\end{minipage}\caption{\textbf{Left}: The Born-Infeld metric $\left(\ref{eq:BIBHf(r)}\right)$
above the dashed black line describes a naked singularity at $r=0$.
Naked singularity solutions in the blue and orange regions have two
photon spheres. In the blue/orange regions, the effective potential
at the inner photon sphere is higher/lower than that at the outer
one. \textbf{Right}: The metric function $f\left(r\right)$ and the
effective potential $V_{\text{eff}}\left(r\right)$ of representative
naked singularity solutions in the blue and orange regions are displayed
in the upper and lower panels, respectively.}
\label{fig:BI}
\end{figure}

Depending on the values of the parameters $a$, $M$, $P$ and $Q$,
the metric function $\left(\ref{eq:BIBHf(r)}\right)$ can describe
a naked singularity at $r=0$ or a black hole. For the black hole
solution with given $a$, $P$ and $Q$, the horizon radius $r_{e}$
and the mass $M_{e}$ of extremal black holes are determined by $f\left(r_{e}\right)=0=d\left(rf\left(r\right)\right)/dr|_{r=r_{e}}$,
which gives
\begin{align}
r_{e} & =\frac{\sqrt{4\left(P^{2}+Q^{2}\right)-a}}{2},\nonumber \\
M_{e} & =\frac{\sqrt{4\left(P^{2}+Q^{2}\right)-a}}{6}+\frac{8\left(P^{2}+Q^{2}\right)\,}{6\sqrt{4\left(P^{2}+Q^{2}\right)-a}}\text{ }_{2}F_{1}\left(\frac{1}{4},\frac{1}{2};\frac{5}{4};-\frac{16a\left(P^{2}+Q^{2}\right)}{\left[4\left(P^{2}+Q^{2}\right)-a\right]^{2}}\right).
\end{align}
Moreover, the above equations show that extremal black holes do not
exist if $a>4\left(P^{2}+Q^{2}\right)$. Therefore, when $a<4\left(P^{2}+Q^{2}\right)$
and $M>M_{e}$, the spacetime is a black hole. For $a>4\left(P^{2}+Q^{2}\right)$,
one has $d\left(rf\left(r\right)\right)/dr>0$ and $\lim\limits _{r\rightarrow0}rf\left(r\right)=4\Gamma\left(\frac{1}{4}\right)\Gamma\left(\frac{5}{4}\right)\left[a\left(P^{2}+Q^{2}\right)\right]^{3/4}/\left(3a\sqrt{\pi}\right)-2M$,
which indicates that a naked singularity appears when
\begin{equation}
M<\frac{2\Gamma\left(\frac{1}{4}\right)\Gamma\left(\frac{5}{4}\right)\left[a\left(P^{2}+Q^{2}\right)\right]^{3/4}}{3a\sqrt{\pi}}.
\end{equation}
In the left panel of FIG. \ref{fig:BI}, the dashed black line denotes
the boundary between black hole and naked singularity solutions in
the $a/M^{2}$-$\sqrt{P^{2}+Q^{2}}/M$ parameter space. Specifically,
solutions above and below the dashed black line correspond to naked
singularities and black holes, respectively.

For black hole solutions, we find that the effective potential $V_{\text{eff}}$
of the effective metric $\left(\ref{eq:BIeffM}\right)$ always has
a single maximum outside the event horizon. Nevertheless, naked singularity
solutions with the absence of the event horizon can possess two photon
spheres and one anti-photon sphere in the effective metric. In the
left panel of FIG. \ref{fig:BI}, the parameter regions where two
photon spheres exist are presented in the $a/M^{2}$-$\sqrt{P^{2}+Q^{2}}/M$
parameter space. In the blue/orange region, the potential peak at
the inner/outer photon sphere is higher than that at the outer/inner
photon sphere. The panels in the right column offer representative
examples in the blue and orange regions.

\subsection{ Bardeen Metric}

To avoid singularities, Bardeen proposed a regular black hole model,
which is described by the metric
\begin{equation}
f\left(r\right)=1-\frac{2Mr^{2}}{\left(r^{2}+P^{2}\right)^{3/2}}.
\end{equation}
Here, $M$ is the mass of the Bardeen metric. In \cite{Ayon-Beato:2000mjt},
it showed that $P$ can be interpreted as the monopole charge of the
magnetic field described by a NLED,
\begin{equation}
\mathcal{L}\left(s\right)=\frac{3M}{2P^{3}}\left(\frac{\sqrt{-2sP^{2}}}{1+\sqrt{-2sP^{2}}}\right)^{5/2}.
\end{equation}
Moreover, the Bardeen metric describes a regular black hole when $P\leq4\sqrt{3}/9$.
For $P>4\sqrt{3}/9$, $f\left(r\right)$ is always positive, which
indicates that the Bardeen metric represents a star with the absence
of a singularity at $r=0$. In addition, the energy density and the
pressures are
\begin{align}
\rho & =-p_{1}=\frac{6MP^{2}}{\left(P^{2}+r^{2}\right)^{5/2}},\nonumber \\
p_{2} & =p_{3}=-\rho+\frac{15MP^{2}r^{2}}{\left(P^{2}+r^{2}\right)^{7/2}},
\end{align}
which indicates that NEC and WEC are satisfied. On the other hand,
SEC and DEC are violated when $r$ is small and large enough, respectively.
Furthermore, the effective metric of photons is
\begin{align}
f_{\text{eff}}\left(r\right) & =h_{\text{eff}}^{-1}\left(r\right)=\frac{16\left(r^{2}+P^{2}\right)^{7/2}}{15r^{6}M}f\left(r\right),\nonumber \\
R_{\text{eff}}\left(r\right) & =\frac{32\left(r^{2}+P^{2}\right)^{9/2}}{15Mr^{4}\left(3r^{2}-4P^{2}\right)}.
\end{align}
In the effective metric of regular Bardeen black holes (i.e., $P\leq4\sqrt{3}/9)$,
there exists only one photon sphere outside the event horizon. On
the other hand, when $4\sqrt{3}/9<P<0.781$, we find that the star
solution can have two photon spheres, and the potential peak at the
outer photon sphere is always higher than that at the inner photon
sphere.

\section{Conclusions}

\label{Sec:Con}

In this paper, we first explored the parameter space of parametrized
black holes and some specific black hole models to find parameter
regions where double photon spheres exist outside the event horizon.
Moreover, the energy conditions, i.e., NEC, WEC, DEC and SEC, were
considered for the black holes with double photon spheres. We then
investigated the existence of double photon spheres in the spacetime
sourced by self-gravitating NLED fields, in which photons propagate
along null geodesics of some effective metric. In summary, we found
that

\begin{itemize}
\item Double photon spheres appear outside the event horizon in certain
parameter regions of parametrized black holes. Moreover, $\left\vert g_{tt}\right\vert $
can monotonically increase outside the event horizon, which is required
by SEC. When $g_{tt}=-g_{rr}^{-1}$, one or more energy conditions
are violated, which implies that double-photon-spheres black holes
satisfying the four energy conditions prefer $g_{tt}g_{rr}\neq-1$.
\item In the EMS model satisfying the four energy conditions, scalarized
RN black holes possess double photon spheres in some parameter regions,
which provides a counterexample of the conjecture proposed in \cite{Cvetic:2016bxi}.
\item Double photon spheres exist in naked singularity solutions sourced
by\ BI electromagnetic fields, which satisfies the four energy conditions.
In addition, the star solution of the Bardeen metric can have double
photon spheres, in which NEC and WEC are respected.
\end{itemize}

Our findings suggest that multiple photon spheres can emerge in some
parameter regions of black holes. Considering significant effects
of extra photon spheres, it is highly desirable to check whether a
given black hole solution can have multiple photon spheres outside
the event horizon. If so, it is also important to obtain parameter
regions where multiple photon spheres exist. To achieve this, we provide
the \textit{Mathematica} code, which can efficiently scan the black
hole parameter space with the help of parallel computing to find mutiple-photon-spheres
parameter regions \cite{Guo:2022mps}. Moreover, the code can also
give parameter regions where the energy conditions are satisfied by
inputting the stress--energy tensor or simply the metric if Einstein's
gravity is assumed. For NLED black holes, calculating the effective
metric of NLED fields with $\mathcal{L}\left(s,p\right)$ is also
included in the code. In this paper, we confined ourselves to Einstein's
gravity. In future studies, looking for black holes with multiple
photon spheres in modified gravity is of equivalent importance. We
hope that the code can facilitate the search procedure.

Interestingly, the anti-photon sphere, which is composed of stable
circular orbits and resides between two photon spheres, play a key
role in the instability of black holes. In \cite{Guo:2021enm}, some
quasinormal modes have been shown to be trapped around the anti-photon
sphere for a long time. The long-lived modes may accumulate near the
anti-photon sphere and eventually develop a non-linear instability,
which could destabilize the background spacetime by the backreaction
\cite{Cardoso:2014sna}. Recently, the authors of \cite{Cunha:2022gde}
used fully non-linear numerical evolutions to confirm that the anti-photon
sphere of ultracompact bosonic stars can trigger the instability.
It is of great interest to study non-linear numerical evolutions of
black holes with multiple photon spheres to address the non-linear
instabilities of long-lived modes.
\begin{acknowledgments}
We are grateful to Yiqian Chen and Qingyu Gan for useful discussions
and valuable comments. This work is supported in part by NSFC (Grant
No. 11875196, 11947225, 12105191, 12275183 and 12275184). Houwen Wu
is supported by the International Visiting Program for Excellent Young
Scholars of Sichuan University.
\end{acknowledgments}

 \bibliographystyle{unsrturl}
\bibliography{ref}

\begin{thebibliography}{10}
\bibitem{Guo:2022mps}
The mathematica notebook is publicly available as ancillary files.
\end{thebibliography}

\end{document}